\documentclass[preprint,12pt,review]{elsarticle}

\usepackage[utf8]{inputenc}
\usepackage{amssymb}
\usepackage{amsmath}
\usepackage{graphicx}
\usepackage{xcolor,hyperref}
\usepackage{tikz}
\usepackage{array}
\usepackage{adjustbox}
\usepackage{diagbox}
\usepackage{makecell}\usepackage[flushleft]{threeparttable}

\hypersetup{colorlinks=true, linkcolor=black, filecolor=black, urlcolor=black, citecolor=blue}

\newcolumntype{R}[2]{%
    >{\adjustbox{angle=#1,lap=\width-(#2)}\bgroup}%
    l%
    <{\egroup}%
}

\journal{Computers in Biology and Medicine}

\begin{document}

\begin{frontmatter}

\title{How is remifentanil dosed without dedicated indicator?}

\author[label1,label4]{Bob Aubouin--Pairault}
\author[label2]{Mazen Alamir}
\author[label5]{Benjamin Meyer}
\author[label5]{Remi Wolf}
\author[label1,label3]{Kaouther Moussa}

\affiliation[label1]{organization={UPHF, CNRS, UMR 8201-LAMIH},
            city={Valenciennes},
            postcode={59313},
            country={France}}
\affiliation[label2]{organization={Univ. Grenoble Alpes, CNRS, Grenoble INP, GIPSA-lab},
            city={Grenoble},
            postcode={38000},
            country={France}}
\affiliation[label3]{organization={INSA Hauts-de-France},
            city={Valenciennes},
            postcode={59313},
            country={France}}
\affiliation[label5]{organization={Univ. Grenoble Alpes, Department of Anaesthesia and Critical Care},
            city={Grenoble},
            postcode={38000},
            country={France}}
\affiliation[label4]{organization={Corresponding author, email: bob.aubouin-pairault@uphf.fr, tel: +336 70 45 64 44}}

\begin{abstract}
This study investigates the paradigm of intraoperative analgesic dosage using a data-driven approach based on retrospective clinical data. Remifentanil, an analgesic widely used during anesthesia, presents a dosing challenge due to the absence of an universally accepted indicator of analgesia. To examine how changes in patient state correlate with adjustments in remifentanil target concentration triggered by the practitioner, we analyzed data from two sources: VitalDB (Seoul, Korea) and PREDIMED (Grenoble, France). Results show that only features derived from arterial pressure are consistently associated with changes in remifentanil targets. This finding is robust across both datasets despite variations in specific thresholds. In particular, increases in remifentanil targets are associated with high or rising arterial pressure over short periods (1–2 minutes), whereas decreases are linked to low, stable, or declining arterial pressure over longer periods (5–7 minutes). By capturing anesthesiologists’ dosing strategies we provide a foundation for the future development of closed-loop control algorithms. Beyond the specific example of remifentanil's change prediction, the proposed feature generation and associated sparse fitting approach can be applied to other domain where human decision can be viewed as sensors interpretation.
\end{abstract}


\begin{highlights}
\item Identify arterial pressure features as key drivers of remifentanil dosing 
\item Link short-term rises to increases and long-term declines to decreases
\item The method enable interpretable models of anesthesiologists’ dosing strategies
\end{highlights}


\end{frontmatter}


\section{Introduction}
Dosage of analgesic drugs, such as remifentanil, during general anesthesia largely relies on the clinical experience of the practitioner. In routine practice, anesthesiologists adjust analgesic delivery according to patient characteristics, the surgical plan, and observable signs of pain during the procedure. The last point, detecting and quantifying pain, remains particularly dependent on individual judgment and has been poorly formalized in the literature \cite{schumacherOpioids2020}. For instance, \cite{leguenAutomatedSedationOutperforms2013} proposed a decision tree using Bispectral Index (BIS), heart rate (HR), and mean arterial pressure (MAP), while \cite{gesztesiUseRemifentanilInfusion1999} suggested an approach relying solely on hemodynamic parameters. Nevertheless, no consensus has been reached to date. The absence of standardized guidelines is especially problematic, as under-dosing of analgesics, in addition to being unethical  to let the patient suffer, has been associated with hemodynamic instability~\cite{ledowskiAnalgesiaNociceptionIndex2014} and increased stress hormone production~\cite{chenCorrelationSurgicalPleth2012}, while over-dosing has been linked to higher post-operative pain~\cite{sabourdinPupillometryguidedIntraoperativeRemifentanil2017, bergmannSurgicalPlethIndexguided2013} and possible hyperalgesia~\cite{fletcherOpioidinducedHyperalgesiaPatients2014,guichardOpioidinducedHyperalgesiaPatients2022}. \medskip 

To address this issue, researchers have been proposing indices, derived from physiological measurements, to assess the pain level felt by patients under anaesthesia. Among the proposed methods, some of them have been commercialized. One could mention the Analgesia Nociception Index (ANI) based on heart rate variability, the Surgical Pleth Index (SPI) based on waveform finger plethysmography, pupil measurement-based methods, q-Nox derived from EEG signals, and Nociception Level (NOL) based on the fusion of plethysmography, skin conductance, skin temperature, and finger movements. The review \cite{funckeValidationInnovativeTechniques2017} particularly reported that ANI, SPI, and pupil-based methods are better detectors of painful stimuli compared to classic monitored signals such as BIS, HR, and MAP. A more recent review~\cite{vansantvlietProgressValidationNociception2024} suggests that changes in ANI and NOL particularly correlate with opioid concentrations in the presence of a noxious stimulus and dosing analgesic drugs in accordance with those indicators might lead to a reduction of post-operative pain. However, none of this commercial indicator have been largely accepted in clinical practice. The cause might lie in the lack of clinical evidence of any benefice for the patient, the lack of ergonomy of some solutions or even the cost of those devices \cite{ledowskiObjectiveMonitoringNociception2019}. \medskip 

Despite these advances, closed-loop control of analgesia is still mostly driven by BIS measurement, even though BIS was originally designed to assess hypnosis rather than analgesia. Particularly, in \cite{funckeValidationInnovativeTechniques2017} the authors found no correlation between BIS variation and noxious stimuli. Nevertheless, remifentanil dosage based on BIS values has been tested clinically for two decades with positive results, see meta-analyses \cite{pasinClosedLoopDeliverySystems2017,brogiClinicalPerformanceSafety2017} or more recent studies \cite{joostenAnestheticManagementUsing2020} for instance. While there is no evidence that those control loops improve patient outcome, they lead to better vital sign stability and less workload for anesthesiologists without worsening patient condition. Only a few closed-loop propositions using dedicated analgesia indicators have been tested clinically, such as the analgoscore (based on HR and MAP variation)~\cite{hemmerlingAnalgoscoreNovelScore2007a} or more recently with the ANI monitor \cite{hureauClinicalEfficacySafety2025a}. The dominance of BIS is likely due to its widespread availability in total intraveinous anesthesia (TIVA), the existence of validated dynamic BIS models for in-silico testing, and well-defined clinical targets (40–60), unlike hemodynamic-based dosing. Recently, new pharmacodynamic models describing the joint effects of propofol and remifentanil on heart rate and arterial pressure have been introduced \cite{suPharmacodynamicMechanismbasedInteraction2023}, potentially enabling control strategies based on more physiologically relevant indicators. \medskip

In this study, we investigate how remifentanil is currently dosed in clinical practice using standard monitoring. We retrospectively analyzed data from two data sources (VitalDB \cite{leeVitalDBHighfidelityMultiparameter2022} and PREDIMED \cite{artemovaPREDIMEDClinicalData2019}), and trained parsimonious machine learning algorithms to predict changes in remifentanil concentration target decided by the anesthesiologist using physiological signal as input. The use of parsimonious algorithms allows us then to provide illustrated explanations of the algorithm’s decision and thereby shed light on the rationale of clinical practice. \medskip

In the literature, the authors of \cite{miyaguchiPredictingAnestheticInfusion2021} have already proposed to predict remifentanil target change using retrospective data. Notice however the work in \cite{miyaguchiPredictingAnestheticInfusion2021} focused on remifentanil target increases and compared different algorithms. The model were interpreted using Shapley values \cite{lundbergUnifiedApproachInterpreting2017}, resulting in a different interpretation per model. We extended this work by considering both increases and decreases of the remifentanil target, and by using two different datasets to assess the robustness of the results. Moreover, thanks to the low number of feature selected by our parcimonious method, a simple interpretation of the models is provided to better understand how anesthesiologists dose remifentanil in practice. \medskip

The remainder of the paper is organized as follows: in Section 2 the methodology is introduced, Section 3 presents the results, and they are discussed in Section 4. Finally, a conclusion and future research directions are given in Section 5. \medskip

\section{Method}
In this section, the data used in the study is first described. Next, the framing of the problem is presented. The machine learning algorithm is then detailed, followed by the metrics used to assess the results and the explanation method used to interpret the model. \medskip

\subsection{Data description}

The data for this study comes from the VitalDB database \cite{leeVitalDBHighfidelityMultiparameter2022} and the PREDIMED clinical data warehouse \cite{artemovaPREDIMEDClinicalData2019}, hereafter referred to as the HG dataset. VitalDB provides high-fidelity multiparameter physiological recordings from 6,389 patients who underwent general anesthesia for non-cardiac surgery in Seoul, Korea, between 2016 and 2017. Signal sampling intervals in this dataset range from one to seven seconds, depending on the signal type. The HG dataset includes data from 589 patients who underwent general anesthesia for cardiac and major surgeries in Grenoble, France, between 2016 and 2022, with physiological signals recorded every 30 seconds. \medskip

In this paper, a subset of patients was selected from each database in order to meet the following criteria:

\begin{itemize}
\item Case duration $>$ 1h30;
\item Patient age $>$ 18 years;
\item Arterial pressure invasively recorded;
\item Use of remifentanil through target controlled infusion pump.
\end{itemize}

In addition, because bolus timing is not available in the VitalDB dataset (only the total amount of drugs injected during the anesthesia), we excluded patients who had received boluses of drugs affecting the hemodynamic and hypnotic systems (propofol, midazolam, fentanyl, ephedrine, phenylephrine, and epinephrine). 290 patients were finally selected from the VitalDB database and 541 patients from the HG database. For the HG dataset, only the period before extracorporeal circulation (ECC) was extracted, as ECC phase was not the purpose of the study. A summary of the demographic description of the two datasets is provided in the supplementary material. \medskip

\subsection{Problem framing}

\begin{figure}[h]
\centering
\includegraphics[width = \textwidth]{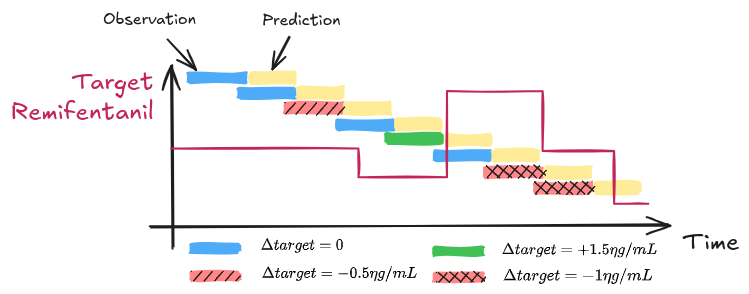}
\caption{Illustration of the framing used to construct the regression problem.}
\label{fig:framing}
\end{figure}

To translate the medical data into an applicable machine learning problem, we used a sliding window framing. This approach has already been used for sepsis risk prediction \cite{lauritsenFramingMachineLearning2021}, yielding realistic results, and was also applied in \cite{miyaguchiPredictingAnestheticInfusion2021} for remifentanil's change prediction. The idea is to divide each window into two parts: the observation and the prediction. Features are extracted from the observation window and fed into the machine learning algorithm to predict a label derived from the prediction window. In this study, the label is the level of change in remifentanil target during the prediction window. This process is illustrated in Figure~\ref{fig:framing}. Window lengths were tuned during model development. \medskip

As mentioned in the introduction, analgesic dosage is mainly based on three criteria: patient characteristics, surgical plan, and signs of nociception detected during the procedure. Since our goal is to characterize the link between vital signs and remifentanil adjustments, we removed changes clearly driven by patient characteristics or surgical plan. Specifically, we excluded the first segment to avoid predicting the initial target, and removed all changes made within five minutes before the first incision, as these are more likely related to surgical planning than patient condition. We also removed segments containing two or more remifentanil target change in the prediction window, to avoid confusing the learning process, and segments with changes of remifentanil target in the observation window, since remifentanil takes a few minutes to affect the hemodynamic system. Finally, for the HG dataset, we removed segments where any bolus of other analgesic was occurring in the observation or prediction window, as these are likely to confound remifentanil dosing.\medskip

The physiological signals considered were BIS, heart rate (HR), and arterial pressure, systolic (SAP), mean (MAP), and diastolic (DAP). The remifentanil target was also included in the observation window. In addition, patient characteristics such as age, body mass index (BMI), and patient health condition with ASA score were added to the feature vector. \medskip

Since increases and decreases of the remifentanil target rely on different physiological patterns, we treated them as two separate machine learning problems. One regressor was trained to predict increases and another to predict decreases. We also treated the two datasets separatly as we expected the dosing strategies to be different in the two hospitals from which the data was recorded. \medskip

\subsection{Machine-learning algorithm}

The machine learning pipeline consisted of four steps: feature extraction, algorithm selection, training, and testing. \medskip

To extract features from the time series in the observation window, we fitted linear regressions and retained slope, and intercept; in addition we added the standard deviation value of regression error for each signal. This resulted in three features per signal and a total of $6\times 3 + 3 = 21$ features, 3 features for 6 signals(BIS, HR, DAP, SAP, MAP, and remifentanil target) and 3 static features (age, BMI, and ASA).  After the extraction, to avoid abnormal features values, we filtered the segments with a z-score (deviation from the mean normalized by the standard deviation) superior to five.\medskip

For the regression algorithm, \cite{miyaguchiPredictingAnestheticInfusion2021} tested six different methods (linear regression, random forest, support vector machine, XGBoost, neural networks, and LSTM) and found no significant differences. To simplify model interpretation, we chose the Least-Angle Regression algorithm with Lasso regularization (LASSOLARS model). The LASSO penalty and it's $L_1$ loss function particularly encourage sparsity in the regression coefficients, leading leads to a parsimonious feature set and facilitate interpretation compared to more dense approaches.  We also considered polynomial augmentation of the feature vector to capture non-linear relationships. \medskip

For training, we used recursive feature elimination, removing the least important features until the performance metric (described in the next subsection) decreased by more than 1\%. We also increased the polynomial order until further augmentation did not improve performance by more than 1\%. An inner three-fold cross-validation was used during this process. \medskip

For evaluation, we used a five-fold cross-validation at the patient level, ensuring independent training and test splits. \medskip

\subsection{Metrics and Explanation}

As the primary metric, we used the area under the receiver operating characteristic curve (AUROC) for the binary problem defined as change vs. no change in the prediction window, following \cite{miyaguchiPredictingAnestheticInfusion2021}. The metric is reported as $mean [min, max]$ in the text and the tables. Although the model is trained as a regression, it can be viewed as a classification task where greater emphasis is placed on significant changes. \medskip

For model explanation, scatter plots of the model outputs in the feature space are shown for the best-performing models across the five-fold cross-validation. This is possible as the number of selected features is reasonable. For a better visualization, points are sorted in the figures such that the ones predicting changes (i.e. maximum value for increases and minimum values for decrease) are highlighted. Moreover, only the lowest model outputs, up to the point where their maximum reaches the 95th percentile of all outputs, are selected to avoid color saturation in the figures. \medskip

\section{Results}
The code to reproduce all the results associated with vitalDB dataset is available here: \url{https://github.com/BobAubouin/Remifentanil_dosage}. Unfortunately, HG data can not be made publicly available due to privacy constraints. \medskip

\begin{figure}[ht]
\centering
\includegraphics[width = 0.8\textwidth]{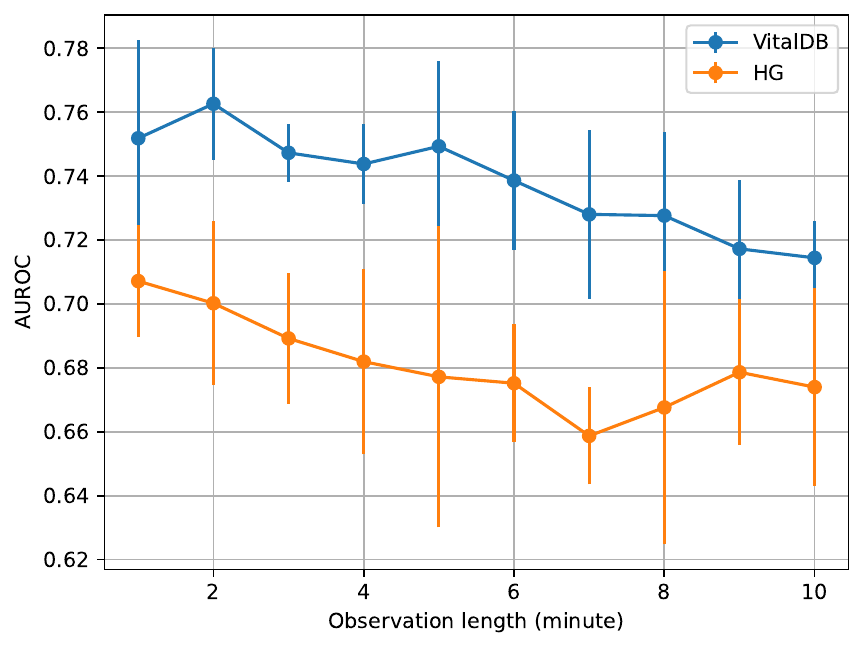}
\caption{Area under the receiver operating characteristic curve (AUROC) for different observation window lengths when predicting increases in remifentanil target on the two datasets. Prediction window length is fixed to 1 minute. Error bars represent the standard deviation across the five folds.}
\label{fig:obs_length_up}
\end{figure}

\begin{figure}[ht]
\centering
\includegraphics[width = 0.8\textwidth]{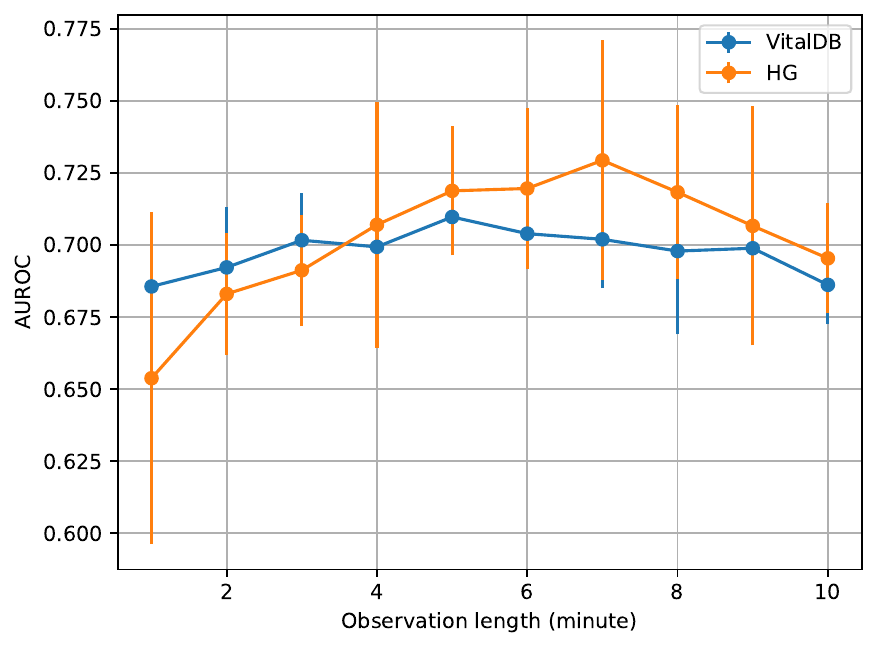}
\caption{Area under the receiver operating characteristic curve (AUROC) for different observation window lengths when predicting decreases in remifentanil target on the two datasets. Prediction window length is fixed to 1 minute. Error bars represent the standard deviation across the five folds.}
\label{fig:obs_length_down}
\end{figure}

In Figures~\ref{fig:obs_length_up} and~\ref{fig:obs_length_down}, the relationship between the length of the observation window and model performance for predicting increases and decreases in remifentanil target is shown. In both datasets, increases in remifentanil target are better predicted with short observation windows of 1-2 minutes, whereas decreases are better predicted with windows of 5–7 minutes. \medskip

\begin{figure}[ht]
\centering
\includegraphics[width = 0.8\textwidth]{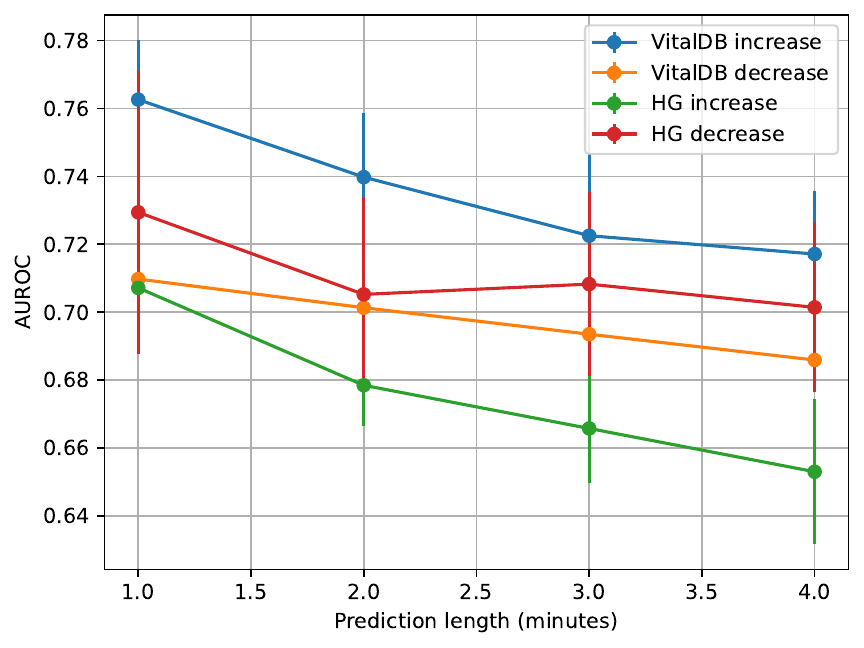}
\caption{Area under the receiver operating characteristic curve (AUROC) for different prediction window lengths using the optimal observation length. Error bars represent the standard deviation across the five folds.}
\label{fig:pred_length}
\end{figure}

Figure~\ref{fig:pred_length} shows the influence of the prediction window length on model performance. In all cases, increasing the prediction window length decreases AUROC. \medskip

Table~\ref{tab:final_results} presents the main results of the predictive models for the best framing parameters. More metrics are also given in supplementary material. Note that the polynomial augmentation of the feature vector was only improving performance for predicting increases of target in the vitalDB dataset with a degree of two. Thus, the other results are simple linear regression models. \medskip

\begin{table}
\centering
\caption{Results of the predictive models for the best framing parameters for each setup. Results are presented as mean [min, max] across the five folds.}

\resizebox{0.9\textwidth}{!}{
\begin{tabular}{l|c|c|c|c}
    \diagbox[innerleftsep=0pt,
    innerrightsep=5.5mm]{\textbf{Setup}}{\textbf{Metric}}
    & \makecell{\textbf{Obs length} \\ \textbf{(minutes)}} 
    & \makecell{\textbf{Prevalence} \\ \textbf{of change}} 
    & \textbf{AUROC} 
    & \makecell{\textbf{Nb of} \\ \textbf{features}} \\
\hline
\textbf{Increase - VitalDB} & 2 & 2.8 \% [2.8\%, 2.9\%] & 76\% [74\%, 78\%]  & 4 [4,4] \\
\textbf{Increase - HG} & 1 & 1.1 \% [1.0\%, 1.1\%] & 70\% [66\%, 72\%]  & 3.2 [3,4] \\
\textbf{Decrease - VitalDB} & 5 & 2.2 \% [2.1\%, 2.2\%] & 71\% [70\%, 73\%]  & 3 [3,3] \\
\textbf{Decrease - HG} & 7 & 2.8 \% [2.8\%, 2.9\%] & 73\% [66\%, 78\%]  & 3.4 [3,5] \\
\end{tabular}}
\label{tab:final_results}
\end{table}

\begin{figure}
    \centering
    \includegraphics[width = 1\textwidth]{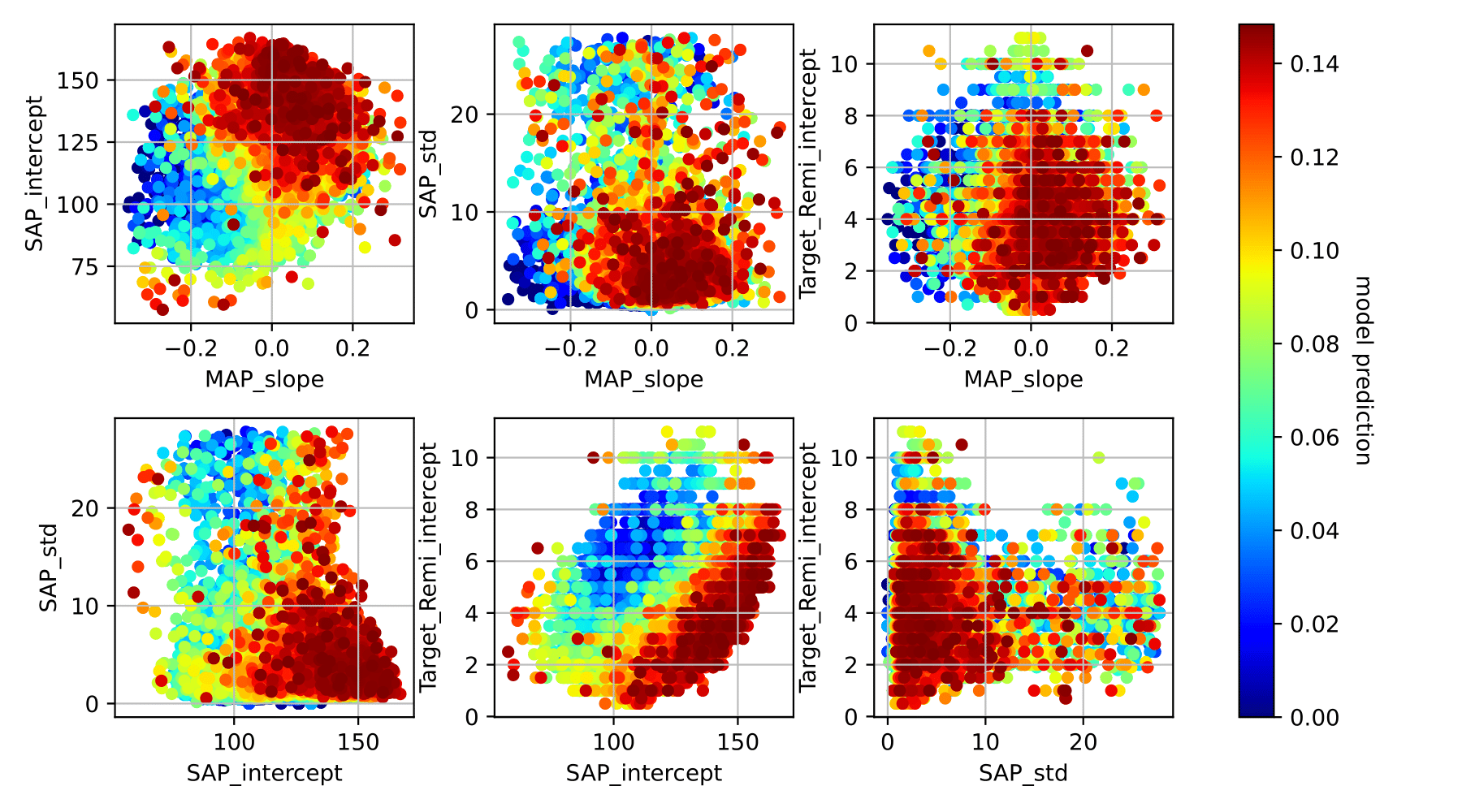}
    \caption{Scatter plot of the model's output for the model trained to predict increase of remifentanil on the VitalDB dataset. Only the lowest model outputs, up to the point where their maximum reaches the 95th percentile of all outputs, are selected to avoid color saturation. }
    \label{fig:scatter_plot_vdb_up}
\end{figure}
\begin{figure}
    \centering
    \includegraphics[width = 1\textwidth]{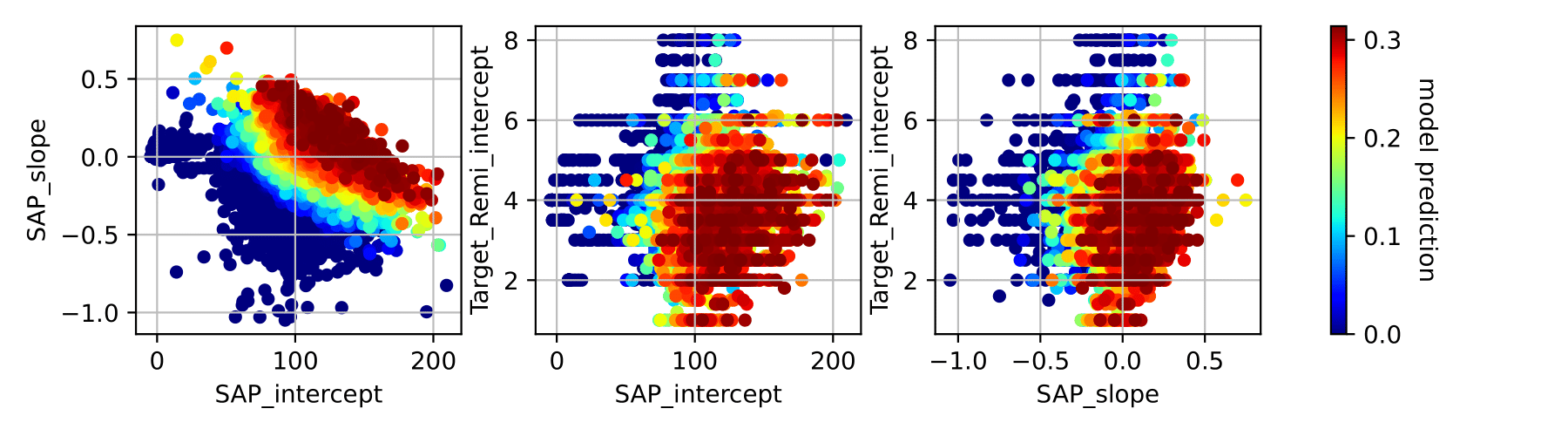}
    \caption{Scatter plot of the model's output for the model trained to predict increase of remifentanil on the HG dataset. Only the lowest model outputs, up to the point where their maximum reaches the 95th percentile of all outputs, are selected to avoid color saturation.}
    \label{fig:scatter_plot_chu_up}
\end{figure}
\begin{figure}
    \centering
    \includegraphics[width = 1\textwidth]{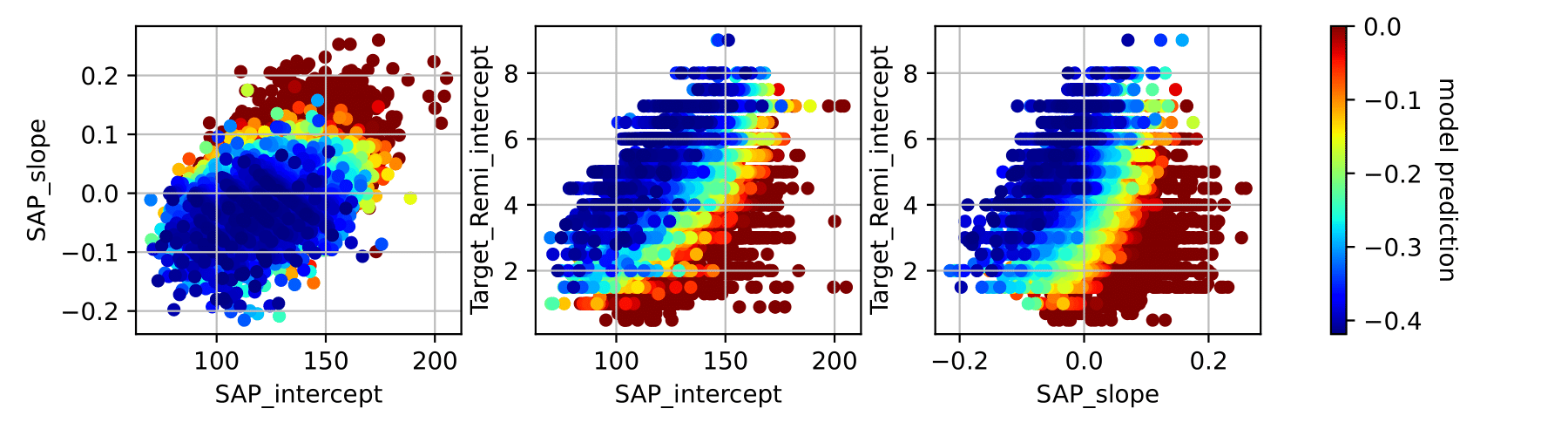}
    \caption{Scatter plot of the model's output for the model trained to predict decrease of remifentanil on the VitalDB dataset. Only the lowest model outputs, up to the point where their maximum reaches the 95th percentile of all outputs, are selected to avoid color saturation.}
    \label{fig:scatter_plot_vdb_down}
\end{figure}
\begin{figure}
    \centering
    \includegraphics[width = 1\textwidth]{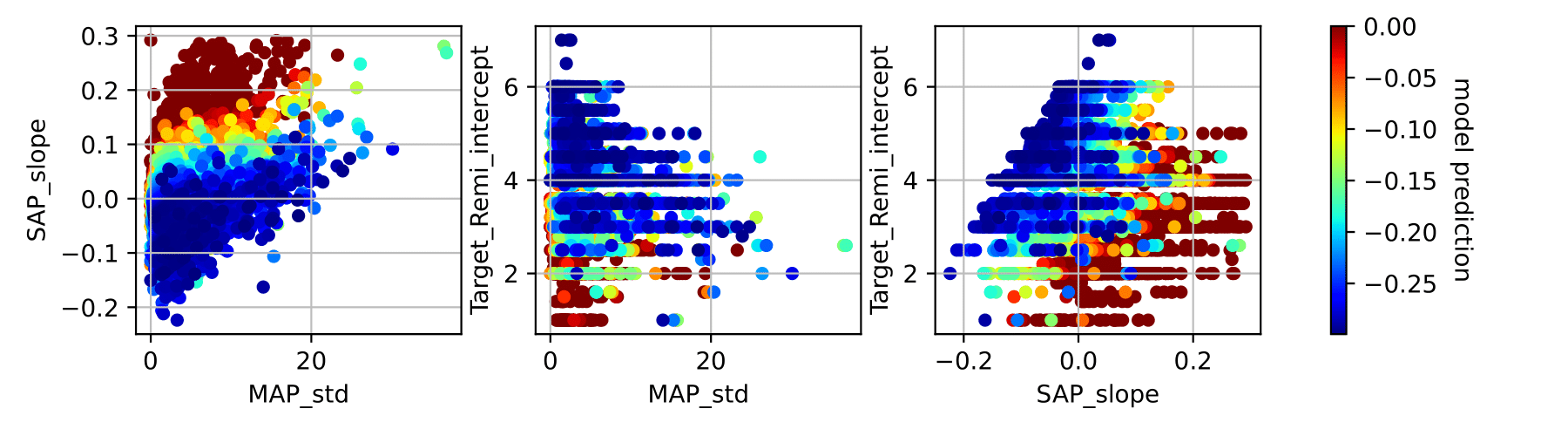}
    \caption{Scatter plot of the model's output for the model trained to predict decrease of remifentanil on the HG dataset. Only the lowest model outputs, up to the point where their maximum reaches the 95th percentile of all outputs, are selected to avoid color saturation.}
    \label{fig:scatter_plot_chu_down}
\end{figure}

For model interpretation, we selected the model trained on the fold associated with the best performance on the validation set. For those models the selected features are given below:
\begin{itemize}
    \item \textbf{Increase - VitalDB}: SAP intercept, SAP slope, SAP error std, Target remifentanil intercept
    \item \textbf{Increase - HG}: SAP intercept, SAP slope, Target remifentanil intercept
    \item \textbf{Decrease - VitalDB}: SAP intercept, SAP slope, Target remifentanil intercept
    \item \textbf{Decrease - HG}: MAP error std, SAP slope, Target remifentanil intercept
\end{itemize}
Figures~\ref{fig:scatter_plot_vdb_up},~\ref{fig:scatter_plot_chu_up},~\ref{fig:scatter_plot_vdb_down} and~\ref{fig:scatter_plot_chu_down} display the scatter plot of the model output in the selected feature space. In those figures MAP stands for Mean Arterial pressure and SAP for Systolic Arterial Pressure. Moreover, intercept, slope and std denote the features extracted from those signals. One can observe from these figures that only features from the arterial pressure and the current target of remifentanil are selected by our method to predict the action of the anesthesiologists.  \medskip

To verify that the proposed feature selection process is statistically meaningful, we provide in Supplementary Material 2 the results of ordinary least squares regression. Specifically, for each feature, we provide the p-value associated with the null hypothesis that its regression coefficient is equal to zero. We observe that the features selected by our methodology correspond to the lowest p-values, indicating that the method has indeed selected the features most strongly correlated with the output. \medskip

\section{Discussion}
The goal of this research was to study remifentanil doing strategies during general anesthesia in order to capture the implicit knowledge of prationners. To this end, we trained parsimonious machine learning algorithms to predict changes in remifentanil target based on physiological signals. \medskip

In the literature, the authors of \cite{miyaguchiPredictingAnestheticInfusion2021} proposed predicting changes in remifentanil target using retrospective data tackling exclusively increases in remifentanil target and compared different algorithms. Blood pressure, remifentanil rate and heart rate were the most important features. Particularly, high arterial pressure value, low remifentanil rate and high heart rate was associated with important Shapley values and thus higher probability of remifentanil increase, which is clinically relevant. In \cite{yuInverseReinforcementLearning2019}, inverse reinforcement learning was applied to infer the reward function of practitioners in order to dose propofol. While this approach is interesting for directly inferring closed-loop algorithms from data, some technical issues remain: the form of the reward function was a priori fixed for each signal, and only the weighting between the costs of each signal was inferred. Moreover, despite discretizing dosing (not changes) into four zones, the resulting algorithm achieved only 54\% agreement between its output and the actual choices of the medical team. \medskip

Different conclusions can be drawn from the results presented in the current article. Regarding the length of the prediction window, our findings are consistent with \cite{miyaguchiPredictingAnestheticInfusion2021}, which reported that predicting remifentanil target is harder as the prediction window length increases. However, with respect to the length of the observation window, our results differ from those of \cite{miyaguchiPredictingAnestheticInfusion2021}, which found no performance difference depending on observation length. In our experiments, short windows (one to two minutes) were better for predicting increases in remifentanil target, whereas longer windows (five to seven minutes) were better for predicting decreases. This discrepancy between increases and decreases in remifentanil target may be explained by the different nature of these actions: increases are usually made in response to painful stimuli, leading to sudden changes in vital signs, whereas decreases are usually made more gradually once surgery has entered a calm phase and the patient shows stable physiological responses. The difference with \cite{miyaguchiPredictingAnestheticInfusion2021} could be due to our feature extraction process, which computes signal trends, while in their study raw time series were used as input to the machine learning algorithms. The computation of signal trends before the use of regressor in our method may have helped to better capture long-term tendancy. \medskip

Looking at model performance, for predicting increases in remifentanil target we found an AUROC of 76\% [74\%, 78\%] on the VitalDB dataset and 71\% [68\%, 73\%] on the HG dataset. Compared to the results of \cite{miyaguchiPredictingAnestheticInfusion2021}, which reported AUROC values of 75.2\% with logistic regression and 75.3\% with LSTM (no confidence interval given), our results are similar despite differences in the datasets, particularly the lower performance on the HG dataset. For predicting decreases in remifentanil target, we found an AUROC of 71\% [70\%, 73\%] on the VitalDB dataset and 73\% [66\%, 78\%] on the HG dataset. To our knowledge, no other study has attempted to predict decreases in remifentanil target. Overall, these results indicate fair performance while also highlighting the difficulty of the task, related to the poor prevalence of target change in the data, leading to an imbalance problem. In particular, given the low precision, it is clear that such a model could not be used directly in a closed-loop system. However, the aim of this study was to understand how anesthesiologists dose remifentanil using TCI pumps. In this regard, the analysis of feature importance and the model's outputs in feature space provides useful insights into anesthesiologists’ decision processes. \medskip

The feature selection procedure enabled us to obtain parsimonious models with a low number of features. Overall, the features selected are extracted only from the arterial pressure signal (MAP or SAP) and the current remifentanil target concentration. In particular, it can be observed that dosing adjustments are made to maintain the remifentanil target concentration within a given range (2–6 ng/mL in the VitalDB dataset and 2.5–5 ng/mL in the HG dataset). Increases in remifentanil target are associated with high or rising arterial pressure in the last minute(s), whereas decreases in remifentanil target are associated with low, stable, or decreasing arterial pressure over a longer period (five to seven minutes). Although decisions are not associated with strict thresholds—since they depend on both practitioner and patient—it is noteworthy that our method yielded consistent results across both datasets. The interpretation of the models for increases is also consistent with the findings of \cite{miyaguchiPredictingAnestheticInfusion2021}, although heart rate was not part of the selected feature with in our methodology. We were also able to link arterial pressure tendency to the probability of increase were the analysis of \cite{miyaguchiPredictingAnestheticInfusion2021} only focused on absolute value interpretation. \medskip

Overall, this study provides valuable insights into how anesthesiologists dose remifentanil using target controlled infusion pumps. These conclusions are in agreement with clinical knowledge that link remifentanil dosage and effect on arterial pressure \cite{schumacherOpioids2020}. it is interesting to observe that simple machine learning algorithms can extract this knowledge from retrospective data and produce robust results across two different datasets. The next step will be to use this knowledge to design control algorithms that mimic anesthesiologists’ behavior, rather than attempting to dose remifentanil optimally according to a predefined notion of optimality, which often diverges from clinical practice. Although such an algorithm will only dose remifentanil reactively to pain, research on preemptive analgesia agree that the timing of analgesic administration can be decoupled from the surgical incisions \cite{brennanPreventiveAnalgesiaReduce2005,roseroPreemptivePreventiveMultimodal2014}.\medskip

This investigation also has limitations. First, both datasets have not been created specifically for this study and originate from only two hospitals. It would be valuable to confirm these results on a dataset designed for this purpose, including annotations of practitioners’ reasoning and surgical timelines. Second, for performance assessment, we chose to dichotomize the problem, which allowed us to compute simple metrics and compare with \cite{miyaguchiPredictingAnestheticInfusion2021}. However, this dichotomization results in a loss of information, since we only evaluate whether the model predicts a change, without accounting for the magnitude of that change. Future work could develop custom metrics to better assess model performance. Finally, we focused on understanding how anesthesiologists dose remifentanil. Given the overlap of multiple tasks and the variable attention of practitioners, it is unlikely that all dosing decisions are homogenous. Noteworthy, in clinical practice, the observed changes decided by the anesthesiologist are generally of small amplitude ($\pm$ 0.1 or 0.2 ng/ml) with a step-by-step approach to adapt to the clinical situation. Thus, directly mimicking practitioners may not be the best approach to designing a control algorithm. Nonetheless, we believe that understanding anesthesiologists’ dosing behavior is a necessary step before attempting to design such algorithms. \medskip

\section{Conclusion}

In this paper, we propose a data-driven approach to understand how anesthesiologists dose remifentanil during TIVA procedures. Using a parsimonious machine learning algorithm on two retrospective datasets, we were able to predict both sided changes in remifentanil targets with fair performance. In particular, the model interpretation, consistent with clinical knowledge, suggests that increases in remifentanil are primarily triggered by high and rising blood pressure over short periods, while decreases are associated with low and stable blood pressure over longer intervals. In the absence of reliable, widespread diffused, and clinically accepted measures of nociception balance, this work provides valuable insights into how closed-loop algorithms could be designed to produce dosage strategies for remifentanil. \medskip

Future work will focus on leveraging the knowledge acquired in this study to design a control algorithm that replicates practitioner behavior. We believe such an algorithm would be more readily accepted by medical teams, as it would align with their reasoning rather than attempting to optimize remifentanil dosing according to a predefined notion of optimality that often diverges from clinical practice. \medskip

\section*{CRediT authorship contribution statement}
\textbf{Bob Aubouin-Pairault}: Data curation, Investigation, Methodology, Software, Visualization, Writing - original draft, Writing - review $\And$ editing; \textbf{Mazen Alamir}: Conceptualization, Investigation, Resources, Supervision, Validation, Writing - review $\And$ editing; \textbf{Benjamin Meyer}: Validation, Writing - review $\And$ editing; \textbf{Remi Wolf}: Validation, Resources, Writing - review $\And$ editing; \textbf{Kaouther Moussa}: Conceptualization, Investigation, Funding acquisition, Project administration, Resources, Supervision, Validation, Writing - review $\And$ editing.

\section*{Declaration of competing interest}
The authors declare that they have no known competing financial interests or personal relationships that could have appeared to influence the work reported in this paper.

\section*{Declaration of generative AI and AI-assisted technologies in the writing process}
During the preparation of this work the authors used ChatGPT in order to improve readability and language. After using this tool, the authors reviewed and edited the content as needed and take full responsibility for the content of the publication. 

\section*{Acknowledgments}
This work was supported by the Clinical project, funded by the ANR (French National Research Agency) under grant ANR-24-CE45-4255.

\bibliographystyle{ieeetr}
\bibliography{bibli}

\newpage

\begin{center}

\Large{Supplementary Material for paper "How is remifentanil dosed without dedicated indicator?"}

\normalsize
Bob Aubouin--Pairault, Mazen Alamir, Benjamin Meyer, Remi Wolf, Kaouther Moussa
\end{center}

\setcounter{section}{0}

\section{Data description}
\begin{table}[h!]
\centering
\caption{Demographic description of the vitalDB datasets.}
\begin{tabular}{lc}
   \hline
 & Median [25th, 75th percentiles] \\
 \hline
Age (years) & 57 [45, 63] \\
Weight (kg) & 62 [55.2, 70] \\
Height (cm) & 165 [157, 173] \\
Duration (h:min) & 4:00 [3:10, 5:05] \\
ASA & 2 [2, 2] \\
Female/Male & 40\% / 60 \% (115 / 175) \\
\textbf{Surgery type} &  \\
   \hspace{5mm} Thoracic & 127 \\
   \hspace{5mm} Digestive & 82 \\
   \hspace{5mm} Vascular & 4 \\
   \hspace{5mm} Other, not cardiac & 77 \\
   \hline
\end{tabular}

\end{table}

\begin{table}[h!]
\centering
\caption{Demographic description of the HG datasets.}
\begin{tabular}{lc}
   \hline
 & Median [25th, 75th percentiles] \\
 \hline
Age (years) & 69 [61, 74] \\
Weight (kg) & 76 [67, 86] \\
Height (cm) & 170 [164, 175] \\
Duration (h:min) & 2:48 [2:20, 3:35] \\
ASA & 3 [3, 3] \\
Female/Male & 26 \% / 74\% (138 / 403) \\
\textbf{Surgery type} &  \\
   \hspace{5mm} Cardiac & 506 \\
   \hspace{5mm} Vascular & 24 \\
   \hspace{5mm} Digestive & 9 \\
   \hspace{5mm} Orthopedic & 1 \\
   \hspace{5mm} Thoracic & 1 \\

   \hline
\end{tabular}

\end{table}

\newpage

\newpage

\section{Other results}

\begin{table}[h!]
\centering
\caption{Other metric measure for the performance of the different models. AUPRC stands for Area Under the Precision-Recall Curve. Sensitivity, Specificity and Precision are computed at the optimal point of the ROC curve (maximal Youden index).}

\resizebox{\textwidth}{!}{
\begin{tabular}{l|c|c|c|c}
\textbf{Setup} & \textbf{AUPRC (\%)} & \textbf{Sensitivity (\%)} & \textbf{Specificity (\%)} & \textbf{Precision (\%)} \\
\hline
\textbf{Increase - VitalDB} & 9.7  [9.3, 9.9] & 69 [63, 80] & 71 [65, 75]  & 6.7 [6.1, 7.4]\\
\textbf{Increase - HG} & 3.1  [2.1, 3.8] & 65 [56, 75] & 70 [57, 80]  & 2.4 [1.7, 2.7]\\
\textbf{Decrease - VitalDB} & 6.3 [5.4, 6.7] & 66 [56, 81] & 67 [56, 76]  & 4.4 [4.0, 5.0]\\
\textbf{Decrease - HG} & 2.1  [1.3, 2.9] & 68 [56, 82] & 71 [52, 77]  & 1.7 [1.2, 2.1]\\
\end{tabular}
}
\end{table}

\bigskip
In the following tables, we report the p-value associated with the results of the ordinary least squares regression to predict increases and decreases in remifentanil target on both datasets. To avoid singularities issues in the regression, we removed MAP and DAP features as they were highly correlated to SAP features. \medskip

\begin{figure}
    \centering
    \includegraphics[width=1\textwidth]{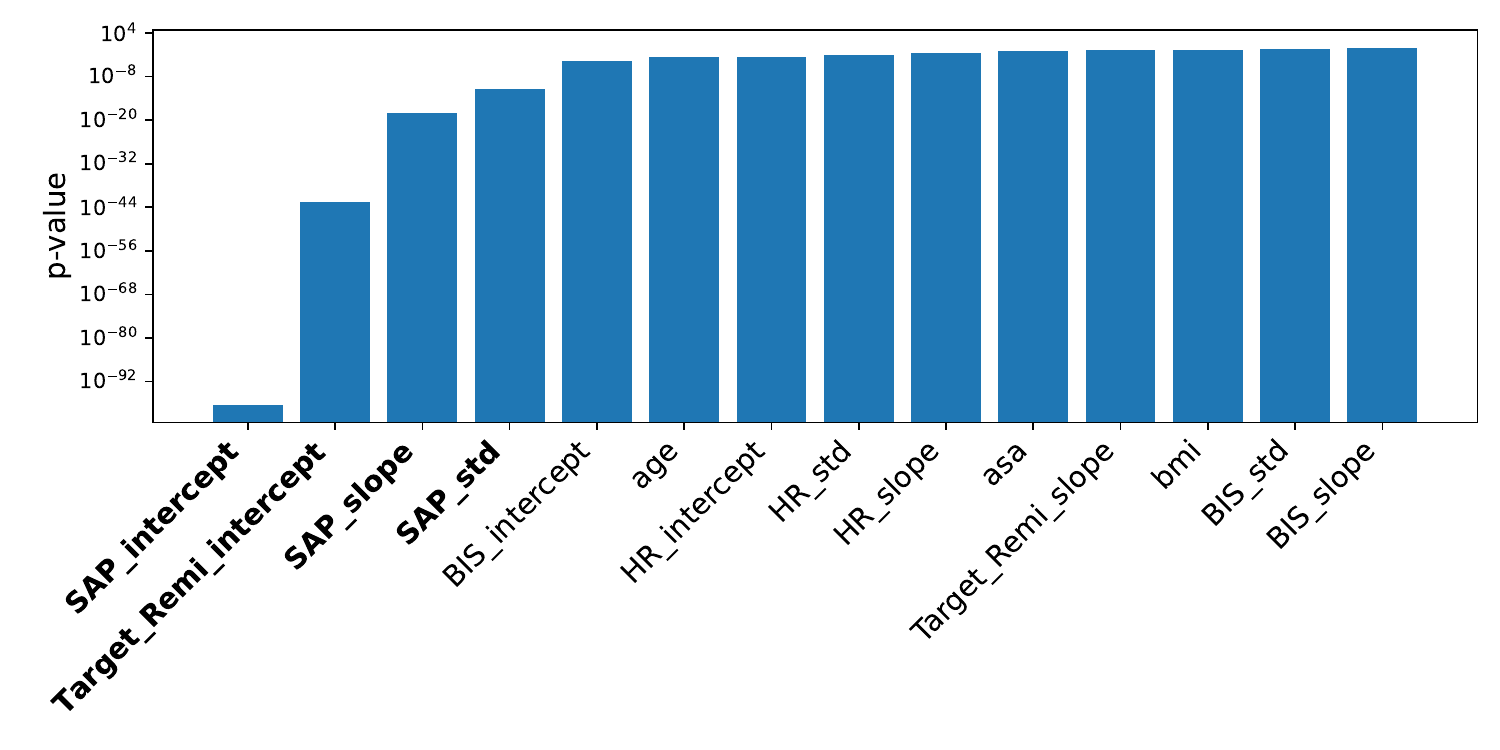}
    \caption{p-value of each feature for the ordinary linear regression to predict increase of remifentanil target in the VitalDB dataset. Bold feature names are the one selected by our methodology. Note that in out methodology 'MAP\_slope' rather that 'SAP\_slope' was the selected feature.}
\end{figure}

\begin{figure}
    \centering
    \includegraphics[width=1\textwidth]{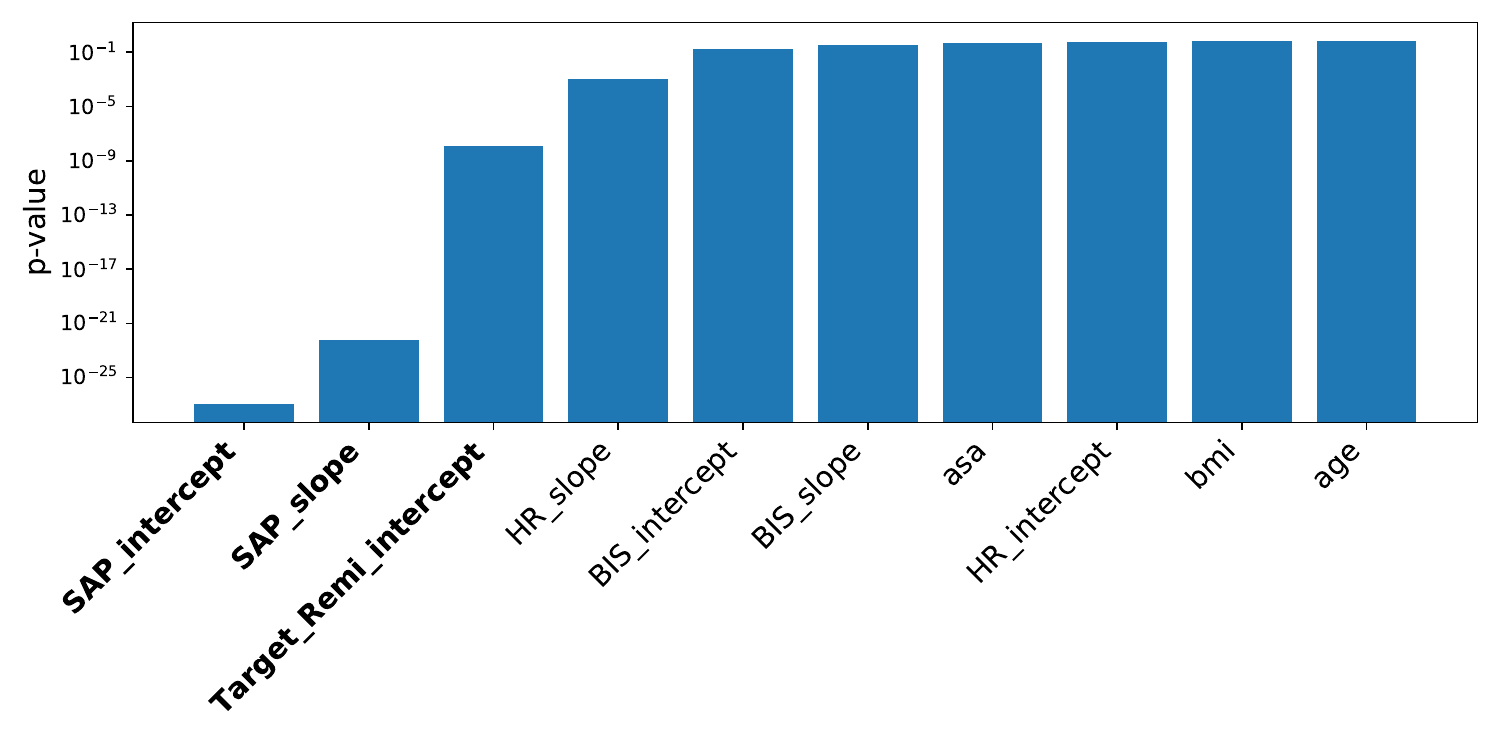}
    \caption{p-value of each feature for the ordinary linear regression to predict increase of remifentanil target in the HG dataset. Bold feature names are the one selected by our methodology.}
\end{figure}

\begin{figure}
    \centering
    \includegraphics[width=1\textwidth]{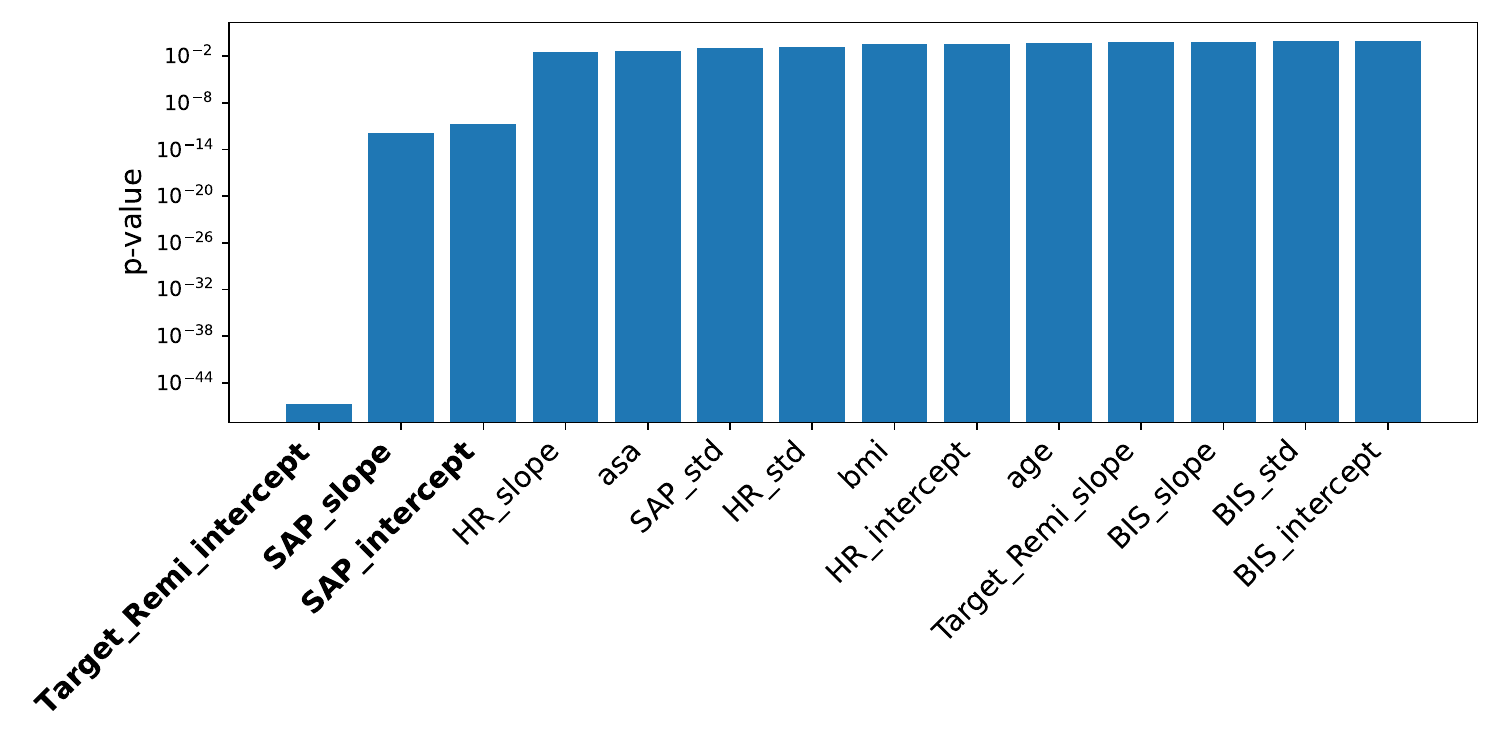}
    \caption{p-value of each feature for the ordinary linear regression to predict decrease of remifentanil target in the VitalDB dataset. Bold feature names are the one selected by our methodology.}
\end{figure}

\begin{figure}
    \centering
    \includegraphics[width=1\textwidth]{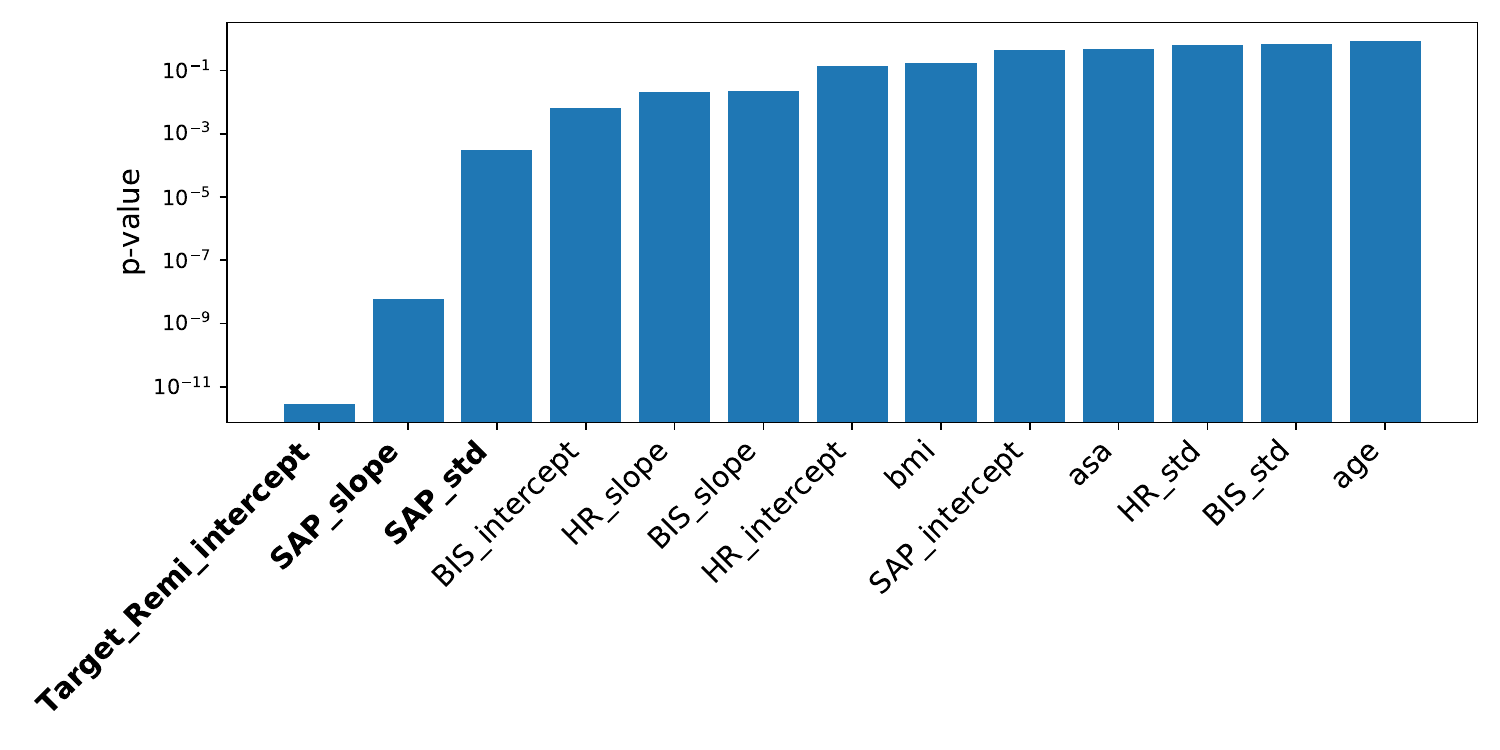}
    \caption{p-value of each feature for the ordinary linear regression to predict decrease of remifentanil target in the HG dataset. Bold feature names are the one selected by our methodology. Note that in out methodology 'MAP\_std' rather that 'SAP\_std' was the selected feature.}
\end{figure}

\end{document}